\documentclass[12pt,draftcls,onecolumn]{IEEEtran}

\usepackage{amsmath}
\usepackage{amssymb}
\usepackage[dvips]{graphicx}
\usepackage{setspace}
\usepackage{epsfig}

\newcommand{\x}{\mathbf{x}}
\newcommand{\y}{\mathbf{y}}

\newcommand{\n}{\mathbf{n}}

\newcommand{\E}{\mathbb{E}}

\newcommand{\tsnr}{{\text{\footnotesize{SNR}}}}

\newcommand{\lit}{\lim_{t\rightarrow \infty}}

\newcommand{\ce}{C_{e}}

\newcommand{\rme}{{\mathrm{e}}}

\newtheorem{theo}{Theorem}
\newtheorem{lemma}{Lemma}

\newtheorem{rem}{Remark}

\date{}

\begin{document}

\title{On the Throughput of Hybrid-ARQ under QoS Constraints}

\author{\authorblockN{Yi Li, M. Cenk Gursoy, and Senem Velipasalar}
\thanks{The authors are with the Department of Electrical Engineering and Computer Science,
Syracuse University, Syracuse, NY 13244. (e-mail:
yli33@syr.edu,
mcgursoy@syr.edu, svelipas@syr.edu).}
}

\maketitle

\begin{spacing}{1.5}

\begin{abstract}
Hybrid Automatic Repeat Request (HARQ) is a high performance
communication protocol, leading to effective use of the wireless channel and the resources with only limited feedback about the channel
state information (CSI) to the transmitter. In this paper, the throughput of HARQ with incremental redundancy (IR) and fixed transmission rate is studied in the presence of quality of service (QoS) constraints imposed as limitations on buffer overflow probabilities. In particular, tools from the theory of renewal processes and stochastic network calculus are employed to characterize the maximum arrival rates that can be supported by the wireless channel when HARQ-IR is adopted. Effective capacity is employed as the throughput metric and a closed-form expression for the effective capacity of HARQ-IR is determined for small values of the QoS exponent. The impact of the fixed transmission rate, QoS constraints, and hard deadline limitations on the throughput is
investigated and comparisons with regular ARQ operation are provided.
\end{abstract}

\begin{IEEEkeywords}
Cumulant generating function, deadline constraints, effective capacity, fading channel, hybrid-ARQ with incremental redundancy (HARQ-IR), QoS constraints, renewal processes.
\end{IEEEkeywords}

\thispagestyle{empty}

\section{Introduction} \label{sec:intro}

Recent years have witnessed a significant growth in the wireless transmission of multimedia content. For instance, as noted in the Cisco Visual Networking Index \cite{Cisco}, mobile video traffic was 51 percent of the mobile data traffic by the end of 2012 and is predicted to grow fast to account for 66 percent of the traffic by 2017. Such multimedia traffic requires certain quality of service (QoS) constraints e.g., in terms of delay, buffer overflow or packet drop/loss probabilities, so that acceptable performance levels can be guaranteed for the end-users. However, satisfying the QoS requirements and providing performance guarantees are challenging in volatile wireless environments in which channel conditions vary over time randomly due to mobility and changing environment. For improved reliability and robustness in wireless transmissions and also efficient resource allocation, one strategy is to adapt the transmissions according to the channel conditions. In particular, if channel state information (CSI) is available at the transmitter, adaptive modulation and coding (AMC) schemes can be employed and transmission power and rate can be varied depending on the channel fading conditions \cite{wirelessbook}.  The required instantaneous CSI can be fed back from the receiver or estimated by the transmitter. With full CSI at the transmitter, AMC achieves higher throughput and/or incurs smaller outage probability \cite{AMC1}.

Automatic Repeat Request (ARQ) is a feedback-based mechanism that can also be used to adapt the wireless transmissions to channel conditions. Different from AMC, ARQ can work with limited CSI at the transmitter. In ARQ schemes, while successfully decoded packets are confirmed with an acknowledgement (ACK) feedback from the receiver, erroneous receptions trigger a negative acknowledgement (NACK) from the receiver and the retransmission of the packet from the transmitter \cite{error_control}. Performance of ARQ protocols has been extensively studied in the literature. In particular, delay/queueing analysis was conducted, for instance in \cite{JGKim}, \cite{LBadia}, and \cite{LBLong}. In \cite{JGKim}, mean delay experienced by a Markovian source over a wireless channel was analyzed when selective-repeat ARQ was employed. Reference \cite{LBadia} investigated the packet delay statistics of the selective-repeat ARQ in Markov channels. A queueing analysis of ARQ protocols together with adaptive modulation and coding strategies was presented in \cite{LBLong} using matrix geometric methods. In \cite{impactqos}, energy efficiency of fixed-rate transmissions was analyzed under statistical queueing constraints when a simple ARQ scheme was employed in outage events.

Combining pure ARQ with error control coding increases the probability of successful transmission and results in the more powerful Hybrid ARQ (HARQ) protocols \cite{HARQ1}. In particular, better adaptation to channel conditions and higher throughput can be achieved by employing HARQ with incremental redundancy (IR). In HARQ-IR, each packet is encoded into a long codeword consisting of a number of subblocks. Initially, the first subblock is transmitted to the receiver. If the packet is decoded correctly using only the first subblock of the codeword, the receiver sends an ACK and the transmitter initiates the transmission of a new packet. In the case of a decoding failure and the reception of a NACK from the receiver, the transmitter sends the next subblock of the codeword, which is jointly decoded at the receiver with the previously received subblock. Therefore, at the receiver, information accumulates with the reception of each subblock until the current packet is decoded successfully. The throughput of HARQ protocols was studied in \cite{throughputharq} from an information-theoretic perspective and it was shown that the throughput of HARQ-IR could approach the ergodic capacity for large transmission rates with only limited CSI. More recently, performance of HARQ in Rayleigh block fading channels was investigated via a mutual information-based analysis in \cite{PWu}, and long-term average rates achieved with HARQ were characterized under constraints on the outage probability and the maximum number of HARQ rounds. In \cite{JChoi}, the tradeoff between energy efficiency and transmission delay in wireless multiuser systems employing HARQ-IR was studied.



In this paper, different from prior work, we analyze the throughput of HARQ-IR protocols in the presence of statistical QoS requirements imposed as constraints on the buffer overflow probability. In particular, we employ the effective capacity formulation \cite{eff_cap} to characterize the maximum constant arrival rates that can be supported by wireless systems employing HARQ-IR protocols while providing statistical QoS guarantees at the same time.

The rest of this paper is organized as follows. In Section \ref{sec:systemmodel}, we introduce the channel model and describe the HARQ-IR scheme. In Section \ref{sec:effectivecap}, we provide our main result on the effective capacity of HARQ-IR in terms of the statistical properties of random transmission time, and discuss the impact of QoS requirements and hard-deadline constraints. Numerical results are given in in Section \ref{sec:numerical} and the paper is concluded in Section \ref{sec:conclusion}. The proof of the main result is relegated to the Appendix.

\section{System Model} \label{sec:systemmodel}
\subsection{Fading Channel}
We consider a point-to-point wireless link and assume that block fading is experienced in the channel. More specifically, in each block of duration $m$ symbols, fading is assumed to stay fixed and then change independently in the subsequent block. In the $i^{th}$ block, the transmitter sends the $m$-dimensional signal vector $\x_i$ with average energy $\E\{\|\x_i\|^2\} = m\mathcal{E}$, and the received signal can be expressed as
\begin{equation}\label{eqsys1}
  \y_i=h_i\x_i+\n_i \quad i=1,2,\ldots
\end{equation}
where $h_i$ is the channel fading coefficient in this block, and $\n_i$ denotes the noise vector with independent and identically distributed (i.i.d.) complex, circularly-symmetric Gaussian components with zero-mean and variance $N_0$. Then, the instantaneous capacity in each fading block can be expressed as
\begin{equation}\label{eqsysec}
  C_i=\log _{2}(1+\tsnr z_i)\quad \text{bit/s/Hz}
\end{equation}
where $\tsnr = \frac{\E\{\|\x\|^2\}}{E\{\|\n\|^2\}} = \frac{m \mathcal{E}}{m N_0} = \frac{\mathcal{E}}{N_0}$ represents the transmitted average signal-to-noise ratio, and $z_i = |h_i|^2$ denotes the magnitude-square of the fading coefficient.

\subsection{HARQ-IR with Fixed-Rate Transmissions}

We assume that the transmitter sends information at the constant rate of $R$ bit/s/Hz and an HARQ-IR protocol is employed for reliable reception. In this scheme, the  messages at the transmitter are encoded according to a certain codebook and the codewords are divided into a number of subblocks of the same length. During each fading block, only one subblock is sent to the receiver. At the receiver side, the transmitted message is decoded according to the current received subblock combined with the previously received subblocks related to the current transmitted message. In this case, information accumulates at the receiver side. According to information-theoretical results \cite{throughputharq}, the receiver can decode the transmitted message at the end of the $M^{\text{th}}$ subblock without error only if $R$ satisfies
\begin{equation}\label{eqharqr}
  R < \sum_{i=1}^{M} \log _{2}(1+\tsnr z_i).
\end{equation}
We assume that the decoder at the receiver has the ability to detect transmission errors reliably. Hence, if $R$ does not satisfy (\ref{eqharqr}), receiver detects the error and sends NACK feedback to the transmitter, triggering the transmission of the next subblock of the same message in the subsequent transmission interval. If, on the other hand, (\ref{eqharqr}) is satisfied, an ACK feedback signal is sent, and the first subblock of a new message is transmitted in the next interval.

We define the random transmission time $T$ of a message as
\begin{equation}\label{eqharqt}
  T=\min \left\{M: R < \sum_{i=1}^{M} \log _{2}(1+\tsnr z_i)\right\}.
\end{equation}
Hence, $T$ denotes the number of block-fading channel uses needed to successfully send a message. It is shown in \cite{throughputharq} that the throughput of this HARQ-IR scheme is given by
\begin{equation}\label{eqharqthp}
  \gamma=\frac{R}{\E\{T\}}=\frac{R}{\mu_1}
\end{equation}
where $\mu_1$ denotes the expected value of $T$.
Additionally, it is proven in \cite{throughputharq} that as $R\rightarrow\infty$, the throughput approaches the ergodic capacity, i.e.,
\begin{equation}\label{eqharqtc}
  \lim_{R\rightarrow\infty} \gamma
  = \E\{\log _{2}(1+\tsnr z_i)\}= \E\{C_i\}.
\end{equation}

\section{Effective Capacity of the HARQ-IR Scheme} \label{sec:effectivecap}

We assume that the transmitter operates under QoS constraints imposed as limitations on the buffer overflow probability. More specifically, we assume that the buffer overflow probability satisfies
\begin{equation}\label{eqqosex}
  \theta=-\lim_{\tau \rightarrow\infty}\frac{\log_e\;P_{r}(Q \ge \tau)}{\tau}
\end{equation}
where $Q$ is the stationary queue length, and $\tau$ is an overflow threshold. The above limiting formulation implies that for sufficiently large threshold $\tau$, we have
\begin{equation}\label{eqqosof}
  P_{r}(Q \ge \tau)\thickapprox e^{-\theta \tau}.
\end{equation}
Hence, the overflow probability decays exponentially fast with rate controlled by the QoS exponent $\theta$. The larger the $\theta$, the smaller the buffer overflow probability becomes. Hence, larger $\theta$ implies that stricter QoS constraints are being imposed.

Throughput under such constraints can be determined through the effective capacity, which provides the maximum constant arrival rates that can be supported while satisfying (\ref{eqqosex}). Effective capacity is  formulated as \cite{eff_cap}
\begin{align}
  \ce&=-\lit\frac{1}{\theta t}\log_e \E\{\rme^{-\theta S_t}\}\\
     &=-\lit\frac{1}{\theta t}\log_e \E\{\rme^{-\theta R N_t }\}\label{eqecec2}
\end{align}
where $S_t$ is the time-accumulated service process representing the total number of bits sent until time $t$.  If we denote the number of successful message transmissions until time $t$ by $N_t$, then $S_t = R N_t$. Note that $N_t$ is the number of renewals made by time $t$ and hence $\{N_t\}$ can be regarded as the renewal counting process with i.i.d. interarrival intervals. More explicitly, we can define $N_t$ as
\begin{gather}
N_t = \max \left\{ k: \sum_{l  = 1}^k T_l < t \right\}
\end{gather}
where $\{T_l\}$ is the i.i.d. sequence of durations of successful transmissions of consecutive messages.
Using the properties of renewal processes, we obtain the following closed-form expression of the effective capacity for small $\theta$ values in terms of the statistical averages of the random transmission time $T$.

\begin{theo} \label{theo:effectivecap}
For the HARQ-IR scheme with fixed rate transmissions, the effective capacity in (\ref{eqecec2}) has the following first order expansion with respect to the QoS exponent $\theta$ around $\theta = 0$:
\begin{align}\label{eqecec3}
\ce &= \frac{R}{\mu_1} - \frac{R^2 \sigma^2}{2\mu_1^3}\theta + o(\theta)
\end{align}
where $R$ is the fixed transmission rate, $\mu_{1}$ and $\sigma^2$ are the mean and variance of the random transmission time $T$, and $\theta$ is the QoS exponent. Note that $\mu_{1}$ and $\sigma^2$ are also functions of $R$. Finally, $o(\theta)$ represents the terms that vanish faster than $\theta$ as $\theta \to 0$, i.e., $\lim_{\theta \to 0} \frac{o(\theta)}{\theta} = 0$.
\end{theo}

\emph{Proof}: See Appendix \ref{sec:proof}.

\begin{rem}
We note that if no QoS constraints are imposed and hence $\theta = 0$, the effective capacity expression in (\ref{eqecec3}) specializes to $\ce = \frac{R}{\mu_1}$ and therefore we recover the throughput formulation obtained in \cite{throughputharq}. Additionally, we notice in (\ref{eqecec3}) that since $R \ge 0$, $\mu_1 \ge 0$, and $\sigma^2 \ge 0$, the introduction of the QoS constraints even with small QoS exponent $\theta$ leads to a loss in the throughput, which was quantified by the term $- \frac{R^2 \sigma^2}{2\mu_1^3}\theta$. Finally, another observation is that while depending only on $\mu_1$ when $\theta = 0$, the throughput starts also depending on the variance, $\sigma^2$, of the random transmission time in the presence of QoS requirements. Indeed, the larger the variance, the smaller the throughput becomes in the regime of small $\theta$.
\end{rem}

\begin{rem}
By the Central Limit Theorem for renewal counting processes \cite{renewal_CLT}, if the inter-renewal intervals have finite variance $\sigma^2$, then we have the following convergence in distribution
\begin{equation*}
  \frac{N_t-\frac{t}{\mu_1}}{\sigma \mu_1^{-\frac{3}{2}} \, t^{1/2}}\longrightarrow \mathcal{N}\left(0,1\right) \quad \text{as } t \rightarrow \infty.
\end{equation*}
Hence, the distribution of $N_t$ tends to a Gaussian distribution with mean $\frac{t}{\mu_1}$ and variance $\frac{\sigma^2 t}{\mu_1^3}$ for large $t$. Now, if we approximate the distribution of $N_t$ as
\begin{align}
f_{N_t} (x) \approx \frac{1}{\sqrt{2\pi \frac{\sigma^2 t}{\mu_1^3}}} \exp\left(-\frac{\left(x - \frac{t}{\mu}\right)^2}{\frac{\sigma^2 t}{\mu_1^3}}\right) \quad \text{for large } t,
\end{align}
then we obtain
\begin{align}
\E\{\rme^{-\theta R N_t }\} \approx \exp\left(- \frac{R}{\mu} \theta t + \frac{R^2 \sigma^2}{2 \mu_1^3} \theta^2 t \right)
\end{align}
which implies that
\begin{align}
\ce=-\lit\frac{1}{\theta t}\log_e \E\{\rme^{-\theta R N_t }\} \approx \frac{R}{\mu_1} - \frac{R^2 \sigma^2}{2\mu_1^3}\theta.
\end{align}
This interesting observation indicates that the characterization in Theorem \ref{theo:effectivecap}, which is valid for small $\theta$, is potentially a good approximation for arbitrary values of $\theta$ as well.
\end{rem}

\subsection*{Hard Deadline Constraints}
Heretofore, we have not considered any restrictions on the random transmission time $T$. Hence, the number of block-fading channel uses needed to successfully send a message can be arbitrarily large especially if the transmission rate $R$ is also large. Indeed, as will be evidenced in the numerical results, throughput improves as $R$ increases but this comes at the cost of increased transmission time. On the other hand, practical systems can require hard deadline constraints for the messages and it is of interest to have bounds on $T$. For instance, we can impose
\begin{gather}
T \le T_u,
\end{gather}
and hence limit the number of HARQ rounds to send a message by $T_u$. More specifically, if $R > \sum_{i=1}^{T_u}\log _{2}(1+\tsnr z_i)$ and hence the message is not correctly decoded at the end of the $T_u^{\text{th}}$ transmission, the message becomes outdated and the transmitter initiates the transmission of the new message.

We can easily see that the characterization in Theorem \ref{theo:effectivecap} applies in the presence of hard-deadline constraints as well, once we adopt the following approach. We define $\hat{T}$ as the total duration of time that has taken to successfully send one message, including the periods of failed transmissions due to imposing the upper bound $T_u$. Now, the probability that $\hat{T} = n + kT_u$, i.e., the probability that the transmission of first $k$ messages have ended in failure due to the deadline constraint and $(k+1)^{\text{th}}$ message is successfully transmitted after $n \le T_u$ HARQ transmissions, can be expressed as
\begin{align}
\Pr\{\hat{T} = n + kT_u\} = \left(\Pr\{T > T_u\}\right)^k \Pr\{T = n\} \quad \text{for } n = 1,2,\ldots,T_u \text{ and } k = 1,2, \dots
\end{align}
where $T$ is as defined in (\ref{eqharqt}). Under the upper bound constraint $T_u$, the new inter-renewal time between successful message transmissions is $\hat{T}$. Hence, only the statistical description of inter-renewal time changes and the throughput formulation in (\ref{eqecec3}) still applies but now with $\mu_1 = \E\{\hat{T}\}$ and $\sigma^2 = \text{var}(\hat{T})$.

%
%
%

Note that the inter-renewal time $\hat{T}$ can grow very fast on average with increasing rate $R$. This is due to the fact that the likelihood to complete the message transmission within $T_u$ intervals becomes small for large $R$. Hence, many message transmissions can fail (i.e., $k$ can become very large) before a successful transmission. More specifically, as $R$ increases, $\Pr\{T > T_u\}$ grows, increasing the probability of large values of $\hat{T}$ and also increasing $\mu_1 = \E\{\hat{T}\}$. This growth is faster than what would be experienced in the absence of hard-deadline constraints and it can lower the throughput significantly if $R$ is larger than a threshold.


\section{Numerical Results} \label{sec:numerical}
In this section, we provide our numerical results. In particular, we focus on the relationship between the transmission rate $R$ and our throughput metric $\ce$. In our results, we both compute the first-order expansion of the effective capacity given in (\ref{eqecec3}) and also simulate the HARQ-IR transmissions and estimate the effective capacity by computing
$
-\frac{1}{\theta t}\log_e \E\{\rme^{-\theta R N_t }\}
$
for large $t$. More specifically, $\E\{\rme^{-\theta R N_t }\}$ is determined via Monte-Carlo simulations. In the numerical analysis, we assume the fading coefficient $h_i$ has a circularly symmetric complex Gaussian distribution with zero mean and variance $1$. Hence, we consider a Rayleigh fading environment.

\begin{figure}
\begin{center}
\includegraphics[width=0.5\textwidth]{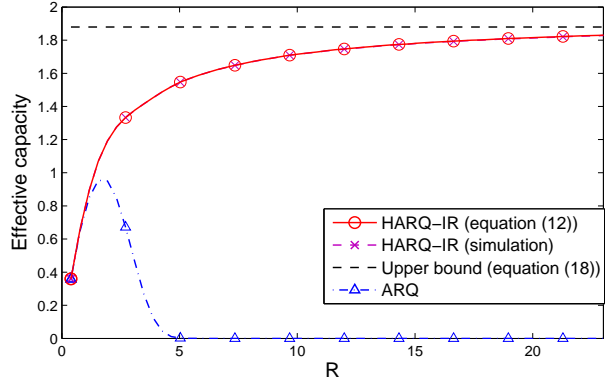}
\caption{Effective capacity $\ce$ vs. transmission rate $R$ at $\text{SNR}=6$ dB and $\theta=0.01$ for both ARQ and HARQ-IR.}\label{fig1}
\end{center}
\end{figure}

\begin{figure}
\begin{center}
\includegraphics[width=0.5\textwidth]{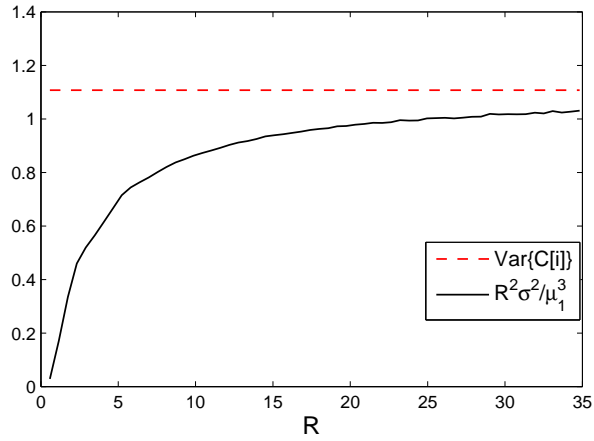}
\caption{$\text{var}(\log_2(1+\tsnr z))$ and $\frac{R^2 \sigma^2}{\mu_1^3}$ vs. transmission rate $R$. $\text{SNR}=6$ dB and $\theta=0.01$.}\label{fig3_2}
\end{center}
\end{figure}

In Fig. \ref{fig1}, we plot the effective capacity $\ce$ as a function of the transmission rate $R$ for both ARQ and HARQ-IR schemes. The throughput of HARQ-IR is plotted both by computing the first-order expansion in (\ref{eqecec3}) and also via simulation. We immediately notice that the effective capacity approximation provided by the first-order expansion is very close to that obtained by simulation for $\theta = 0.01$. Hence, as predicted in Section \ref{sec:effectivecap}, first-order expansion gives an accurate characterization of the throughput of HARQ-IR.
In the figure, we further observe that HARQ-IR significantly outperforms ARQ. Throughput of ARQ initially increases and reaches its peak value at an optimal value $R^*$ beyond which it starts to diminish. Hence, in ARQ, rates higher than the optimal $R^*$ are leading to a large number of retransmissions and resulting in lower throughput. On the other hand, the throughput of HARQ-IR interestingly improves with increasing $R$ and approaches
\begin{equation}
  C_{e, \text{perfect CSI}}=-\frac{1}{\theta}\log_e \E\{e^{-\theta C}\} = -\frac{1}{\theta}\log_e \E\{e^{-\theta \log_2(1+\tsnr z)}\}
\end{equation}
which is the effective capacity of a system in which the transmitter knows the channel fading coefficients perfectly and transmits the data at the time-varying rate of $\log_2(1+\tsnr z)$ in each block. Note that this observation can be seen as the extension of (\ref{eqharqtc}) to the case with QoS constraints. Furthermore, it can be easily verified that the first-order expansion of $C_{e, \text{perfect CSI}}$ is given by
\begin{gather}
C_{e, \text{perfect CSI}} = \E\{\log_2(1+\tsnr z)\} - \text{var}(\log_2(1+\tsnr z)) \frac{\theta}{2} + o(\theta)
\end{gather}
where $\text{var}(\log_2(1+\tsnr z))$ denotes the variance of $\log_2(1+\tsnr z)$. Comparing this expansion with (\ref{eqecec3}) and noting the limiting result in (\ref{eqharqtc}) and the observation in Fig. \ref{fig1}, we expect that $\frac{R^2 \sigma^2}{\mu_1^3}$ approaches $\text{var}(\log_2(1+\tsnr z))$ as $R$ increases, which is verified numerically in Fig. \ref{fig3_2}.



\begin{figure}
\begin{center}
\includegraphics[width=0.5\textwidth]{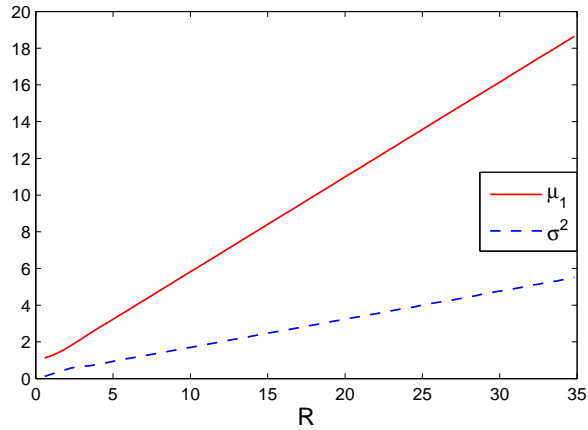}
\caption{Mean $\mu_1$ and the variance $\sigma^2$ of the transmission time $T$ vs. transmission rate $R$. $\text{SNR}=6$ dB and $\theta=0.01$. }\label{fig3}
\end{center}
\end{figure}

The improvement in the throughput of HARQ-IR with increasing $R$ comes at the cost of increased transmission time. This is demonstrated in Fig. \ref{fig3} which shows that both the mean $\mu_1 = \E\{T\}$ and the variance $\sigma^2 = \text{var}(T)$ of the random transmission time $T$ increases with increasing $R$. It is interesting to note that this increased transmission time in HARQ-IR does not have detrimental impact on the throughput under QoS constraints, which is a testament to the efficient utilization of the channel and resources by HARQ-IR. Indeed, it takes more time to send the data but proportionally a large amount of data is sent successfully with HARQ-IR over this extended period of time. Another observation in Fig. \ref{fig3} is at the other end of the line. As $R$ diminishes, $\mu_1$ and $\sigma^2$ approach $1$ and $0$, respectively. This implies from (\ref{eqecec3}) that $\ce \approx R $ for very small $R$, explaining the linear growth of the effective capacity curve of HARQ-IR for small $R$ values in Fig. \ref{fig1}.



\begin{figure}
\begin{center}
\includegraphics[width=0.5\textwidth]{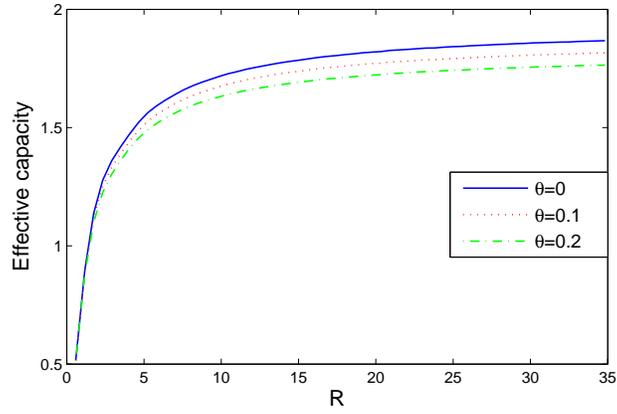}
\caption{Effective capacity $\ce$ of HARQ-IR vs. transmission rate $R$ at $\text{SNR}=6$ dB for different $\theta$ values.}\label{fig2}
\end{center}
\end{figure}
\begin{figure}
\begin{center}
\includegraphics[width=0.5\textwidth]{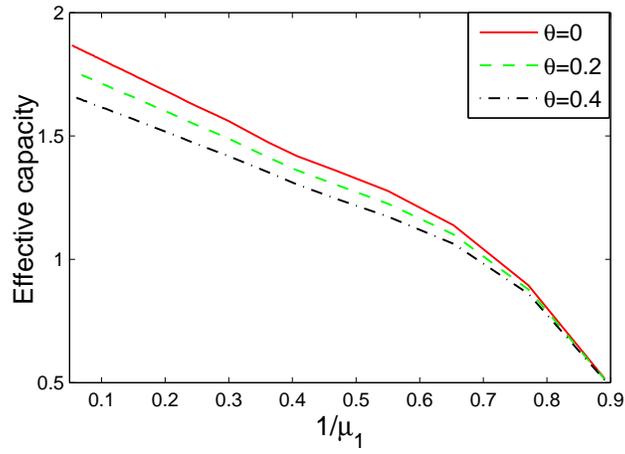}
\caption{Effective capacity $\ce$ vs. $\frac{1}{\mu_1}$ at $\text{SNR}=6$ dB for different $\theta$ values.}\label{fig5}
\end{center}
\end{figure}

In Fig. \ref{fig2}, we plot the effective capacity vs. $R$ curve for different values of the QoS exponent $\theta$. We see that larger $\theta$ values (and hence stricter QoS constraints) expectedly lead to lower throughput. Equivalently, as $\theta$ increases, the same effective capacity is achieved by transmitting at higher rates $R$ and hence by potentially experiencing larger transmission time as depicted in Fig. \ref{fig5}. We note in Fig. \ref{fig5} that especially for high effective capacities, when $\theta$ is increased, the same effective capacity is achieved at smaller values of $\frac{1}{\mu_1} = \frac{1}{\E\{T\}}$.



\begin{figure}
\begin{center}
\includegraphics[width=0.5\textwidth]{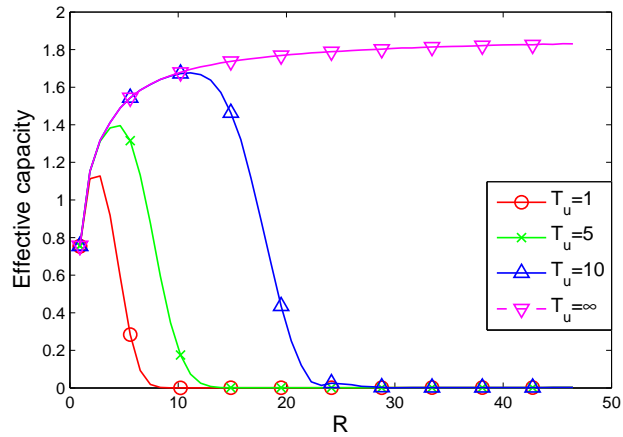}
\caption{Effective capacity $\ce$ vs. transmission rate $R$ for different hard deadline constraints. $\text{SNR}=6$ dB and $\theta = 0.1$}\label{fig6}
\end{center}
\end{figure}

Finally, we address the impact of hard-deadline constraints in Fig. \ref{fig6}. We plot $\ce$ vs. $R$ curves for different values of the upper bound $T_u$ on the transmission time $T$ (or equivalently the number of HARQ rounds). We readily observe that when hard deadline constraints are imposed, there exists an optimal transmission rate $R^*(T_u)$ at which the throughput is maximized and beyond which the throughput starts diminishing. The optimal $R^*(T_u)$ and the achieved maximum throughput get larger for larger $T_u$ while the throughput monotonically increases with increasing $R$ when no deadline constraints are imposed, i.e., when $T_u = \infty$.

\section{Conclusion} \label{sec:conclusion}
In this paper, we have investigated the throughput of HARQ-IR in the presence of QoS constraints imposed as limitations on buffer overflow probabilities. Using the statistical properties of the renewal counting process, we have identified the first-order expansion of the effective capacity of HARQ-IR in terms of the QoS exponent $\theta$. We have shown that the loss in throughput is proportional to $\frac{R^2 \sigma^2}{2\mu_1^3}\theta$ for small $\theta$. We have taken into account hard deadline constraints by imposing an upper bound on the number of HARQ rounds to send a message. We have discussed that the main result on the first-order expansion of the effective capacity holds in the presence of deadline constraints with a modified description of the transmission time. Through numerical results, we have demonstrated that increasing the transmission rate $R$ improves the throughput monotonically in HARQ-IR and makes it approach the throughput of a system with perfect CSI at the transmitter while it initially improves and then lowers the throughput in ARQ. We have noted the superiority of HARQ-IR over the simple retransmission strategy of ARQ. We have also observed that increased throughput with larger $R$ comes at the expense of longer transmission time or equivalently larger number of HARQ-IR rounds. We have shown that the throughput degrades when stricter QoS constraints or hard-deadline constraints are imposed. In particular, we have demonstrated that monotonic growth in the throughput with increasing $R$ is not experienced in the presence of deadline limitations.


\appendix

\subsection{Proof of Theorem \ref{theo:effectivecap}} \label{sec:proof}
The proof of Theorem \ref{theo:effectivecap} is based on the Taylor expansion of the cumulant generating function, which expresses $\ce$ as a polynomial function of $\theta$. After deriving the zeroth and first order coefficients of this polynomial, a closed-form expression for the first-order expansion is obtained for small $\theta$. Before finding the polynomial approximation, we need to show that the moments of the random transmission time $T$ are finite.

Let us denote the $j^{\text{th}}$ moment of the random transmission time $T$ by
\begin{equation}\label{mu}
  \mu_j=\E\{T^{j}\}.
\end{equation}
The following characterization shows that $T$ has finite support for any fixed transmission rate and therefore we have $\mu_j < \infty$ for all $1 \le j < \infty$.

\begin{lemma}\label{theo:moment}
If the expected value of the instantaneous capacity is strictly greater than zero, then for any fixed transmission rate $R$, the random transmission time $T$ has finite support. Hence, all of its moments are finite.
\end{lemma}
\emph{Proof}:
Since $\{z_i\}$ is a sequence of i.i.d. random variables, by the strong law of large numbers \cite[Section 7.4]{book}, we have $\frac{1}{n} \sum_{i=1}^{n} \log_{2}(1+\tsnr z_i)$ converge to $\E\{\log_{2}(1+\tsnr z_i)\} = \E\{C_i\}$ almost surely, i.e., we have
\begin{equation}\label{largenumber}
  \Pr\bigg(\lim_{n\rightarrow \infty} \frac{1}{n} \sum_{i=1}^{n} \log_{2}(1+\tsnr z_i)=\E\{C_i\}\bigg)=1.
\end{equation}
This almost sure convergence implies that \emph{with probability one}, for any given $\varepsilon >0$, there exists a positive integer $n_1$ such that for all $n\geq n_1$
\begin{equation}\label{largenumber1}
  \bigg|\frac{1}{n} \sum_{i=1}^{n} \log_{2}(1+\tsnr z_i)-\E\{C_i\}\bigg|\leq\varepsilon.
\end{equation}
or equivalently
\begin{equation}\label{largenumber1}
  \E\{C_i\} - \varepsilon \leq \frac{1}{n} \sum_{i=1}^{n} \log_{2}(1+\tsnr z_i)\leq \E\{C_i\} + \varepsilon.
\end{equation}
Hence, under the assumption that $\E\{C_i\} > 0$,  we have the following lower bound with probability one for some $0< \varepsilon < \E\{C_i\}$:
\begin{equation}\label{largenumber2}
  \frac{1}{n} \sum_{i=1}^{n} \log_{2}(1+\tsnr z_i)\geq \E\{C_i\}-\varepsilon >0.
\end{equation}
Next, we consider a bound on $\frac{R}{n}$. For a fixed transmission rate $R$, we have
\begin{equation}\label{largenumber3}
  \lim_{n\rightarrow\infty} \frac{R}{n}=0.
\end{equation}
Therefore, for any $\varepsilon_2>0$, there exists an integer $n_2\geq n_1$ such that for all $n \geq n_2$, we have
\begin{equation}\label{largenumber4}
  \frac{R}{n}\leq\varepsilon_2.
\end{equation}
Choosing $\varepsilon_2=E\{C_i\}-\varepsilon$ and using the bound in (\ref{largenumber2}), we have for all $n\geqslant n_2$ that
\begin{align}\label{largenumbe5}
  \frac{R}{n}&\leq \frac{1}{n} \sum_{i=1}^{n} \log_{2}(1+\tsnr z_i)\\
  \intertext{or equivalently}
            R&\leq \sum_{i=1}^{n} \log_{2}(1+\tsnr z_i)
\end{align}
with probability one for all $n \ge n_2$. This implies that the random transmission time $T$ for reliably sending $R$ bits is upper bounded by $n_2$ with probability one, i.e.,  $\Pr(T\leq  n_2)= 1.$

Hence, for any given fixed transmission rate $R$, $T$ has finite support as claimed in the lemma. Hence, the moments $\mu_j = \E\{T^j\} \le n_2^j < \infty$ are finite for all $ 1\le j < \infty$.  \hfill $\square$

Having shown the finiteness of all moments of $T$, we next consider the cumulant generating function of $N_t$, which is the logarithm of the moment generating function of $N_t$, i.e.,
\begin{align}
g(z) = \log \E\{\rme^{z N_t}\}.
\end{align}
This cumulant generating function can be expressed as
\begin{align}
g(z) = \sum_{j = 1}^\infty \kappa_j(t) \frac{z^j}{j !}
\end{align}
where $\kappa_j(t)$ is the $j^{\text{th}}$ order cumulant of $N_t$. Examining (\ref{eqecec2}), we notice that effective capacity is
proportional to the cumulant generating function of $N_t$ and we can write
\begin{align}
\frac{1}{\theta t} \log_e \E\{\rme^{-\theta R N_t}\} &= \frac{1}{\theta t} \sum_{j = 1}^\infty \kappa_j(t) \frac{(-\theta R)^j}{j !}
=\sum_{j = 1}^\infty \frac{\kappa_j(t)}{t} \, \frac{(-1)^j R^j}{j !} \, \theta^{j-1}.
\end{align}
Now the effective capacity can be expressed as
\begin{align}
\ce&=-\lit\frac{1}{\theta t}\log_e \E\{\rme^{-\theta R N_t }\}
\\
&= -\lit \sum_{j = 1}^\infty \frac{\kappa_j(t)}{t} \, \frac{(-1)^j R^j}{j !} \, \theta^{j-1}
\\
&= \sum_{j = 1}^\infty \left(\lit \frac{\kappa_j(t)}{t}\right) \, \frac{(-1)^{j+1} R^j}{j !} \, \theta^{j-1}
\end{align}
It has been proven in \cite{cumulant} that if the moments of $T$ are finite, then the $j^{\text{th}}$ cumulant of $N_t$ can be written as
\begin{align}
\kappa_j(t) = a_j t + b_j + o(1)
\end{align}
for some constants $a_j$ and $b_j$ which depend on the moments of $T$. From this result, we conclude that
\begin{gather}
\lit \frac{\kappa_j(t)}{t} = a_j
\intertext{and hence}
\ce = \sum_{j = 1}^\infty a_j \, \frac{(-1)^{j+1} R^j}{j !} \, \theta^{j-1}. \label{eq:Ce-aj}
\end{gather}
Furthermore, it has been shown in \cite{cumulant} and \cite{renewal} that
\begin{gather}
a_1 = \frac{1}{\mu_1} \quad \text{and} \quad a_2 = \frac{\mu_2 - \mu_1^2}{\mu_1^3} = \frac{\sigma^2}{\mu_1^3}
\end{gather}
where $\mu_1  = \E\{T\}$ and $\mu_2 = \E\{T^2\}$ are the first and second moments of $T$ and $\sigma^2$ is the variance of $T$. Plugging in these values into (\ref{eq:Ce-aj}), we readily obtain
\begin{gather}
\ce = \frac{R}{\mu_1} - \frac{R^2 \sigma^2}{2\mu_1^3}\theta + o(\theta)
\end{gather}
where $o(\theta)$ denote the terms which decay faster than $\theta$, i.e., $\lim_{\theta \to 0} \frac{o(\theta)}{\theta} = 0$. Hence, the desired characterization in Theorem \ref{theo:effectivecap} is proved. \hfill $\square$

\bibliographystyle{ieeetr}
\bibliography{eff_c_ref}

\end{spacing}

\newpage

\end{document}